\documentclass [12pt]{iopart}
\expandafter\let\csname equation*\endcsname\relax
\expandafter\let\csname endequation*\endcsname\relax

\usepackage{amsmath,color,iopams,graphicx,harvard,bm}

\begin{document}
\title[]{Flow in a Circular Expansion Pipe Flow: Effect of a Vortex Perturbation on Localized Turbulence}
\author{Kamal Selvam$^1$, Jorge Peixinho$^1$ and Ashley P. Willis$^2$}
\address{$^1$ Laboratoire Ondes Millieux Complexes, CNRS \& Universit\'e du Havre, 76600 Le Havre, France}
\address{$^2$ School of Mathematics and Statistics, University of Sheffield, Sheffield S3 7RH, UK}
\ead{jorge.peixinho@univ-lehavre.fr}

\begin{abstract}
We report the results of three-dimensional direct numerical simulations for incompressible viscous fluid in a circular pipe flow with a sudden expansion. At the inlet, a parabolic velocity profile is applied together with a finite amplitude perturbation in the form of a vortex with its axis parallel to the axis of the pipe. At sufficiently high Reynolds numbers the recirculation region breaks into a turbulent patch that changes position axially depending on the strength of the perturbation. This vortex perturbation is believed to produce a less abrupt transition than in previous studies with a tilt perturbation, as the localized turbulence is observed via the formation of a wavy structure at a low order azimuthal mode, which resembles an optimally amplified perturbation. For higher amplitude, the localized turbulence remains at a constant axial position. It is further investigated using proper orthogonal decomposition, which indicates that the centre region close to the expansion is highly energetic.
\end{abstract}

\section{Introduction}

The flow through an axisymmetric expansion in a circular pipe is of both fundamental and practical interest. The geometry arises in many applications, ranging from engineering to physiological problems such as the flow past stenoses \cite{Varghese2007}. The bifurcations of flow patterns in sudden expansions have been studied experimentally \cite{Sreenivasan1983,Latornell1986,Hammad1999,Mullin2009} and numerically \cite{Sanmiguel-Rojas2010,Sanmiguel-Rojas2012}.  In all these studies, flow separation after the expansion and reattachment downstream leads to the formation of a recirculation region near the wall.  Its extent grows linearly as the flow velocity increases. 

Numerical simulations and experimental results have shown that the recirculation region breaks axisymmetry once a critical Reynolds number is exceeded. Here, the Reynolds number is defined $Re=Ud/\nu$, where $U$ is the inlet bulk flow velocity, $d$ is the inlet diameter and $\nu$ is the kinematic viscosity. In experiments, the recirculation region loses symmetry at $Re\simeq1139$ \cite{Mullin2009} and forms localized turbulent patches that appears to remain in at a fix axial position \cite{Sanmiguel-Rojas2012,Peixinho2013,Selvam2015}.

Global stability analysis \cite{Sanmiguel-Rojas2010} suggests that the symmetry breaking occurs at a much larger critical $Re$. The reason for the early occurrence of transition in experiments is believed to be due to imperfections, which are very sensitive to the type or the form of the imperfections. These imperfections are modelled in numerical simulations by adding arbitrary perturbations. Small disturbances are likely to be amplified due to the convective instability mechanism, and appear to be necessary to realise time-dependent solutions. Numerical results \cite{Cantwell2010}, have also shown that small perturbations are amplified by transient growth in the sudden expansion for $Re\leq1200$, advect downstream then decay. Simulations in relatively long computational domains, which  accommodate the recirculation region with an applied finite amplitude perturbation at the inlet \cite{Sanmiguel-Rojas2012,Selvam2015}, found the transition to turbulence to occur at $Re\gtrsim1500$, depending upon the amplitude of the perturbation. 

The most basic perturbation is to mimic a small tilt at the inlet, via a uniform cross-flow, on top of the Hagen-Poiseuille flow \cite{Sanmiguel-Rojas2012,Selvam2015,Duguet2015}. This perturbation creates an asymmetry in the recirculation region downstream, which oscillates due to Kelvin-Helmholtz instability, similar to that of a wake behind axisymmetric bluff bodies \cite{Bobinski2014}. At higher $Re$, the recirculation breaks to form localized turbulence. Another possibility is to include a rotation of the inlet pipe, and numerical simulations with a swirl boundary condition \cite{Sanmiguel-Rojas2008}, have shown the existence of three-dimensional instabilities above a critical swirl velocity. Experimental studies have also been conducted \cite{Miranda-Barea2015}, for expansion ratio of 1:8, confirming the existence of convective and absolute instabilities, and also time-dependent states. The higher the $Re$, the smaller is the swirl sufficient for the transition between states to take place. In the present investigation, a small localized vortex perturbation is added at the inlet, without wall rotation, along with the Hagen-Poiseuille flow. This vortex perturbation has been implemented to observe a less abrupt transition to localized turbulence than observed for the tilt case, enabling study of the most energetic modes during the transition.

The goal of the present investigation is to numerically model the expansion pipe flow with a localized vortex perturbation added to the system. In the part 2, the numerical method is presented. Next, in the part 3, the vortex perturbation is described together with the results for the spatio-temporal dynamics of the turbulent patch and the analysis of the localized turbulent patch using proper orthogonal decomposition. Finally, the conclusions are stated in part 4. \\

\section{Numerical method}

Equations governing the flow are unsteady three-dimensional incompressible Navier-Stokes equation for a viscous Newtonian fluid:
\begin{align}
\nabla\cdot \textit{\textbf{v}} &= 0 \label{eqn:conti}\\ 
\dfrac{\partial \textit{\textbf{v}}}{\partial \textit{t}} + 
\textit{\textbf{v}}\cdot\nabla \textit{\textbf{v}}
&= - \nabla P + \dfrac{1}{Re}\nabla^{2} \textit{\textbf{v}} \, ,
\label{eqn:navier}
\end{align}
where $\bm{v}=(u, v, w)$ and $P$ denote the scaled velocity vector and pressure 
respectively. The equations \eqref{eqn:conti} and \eqref{eqn:navier} were non-dimensionalised using inlet $d$ and $U$. The time scale and the pressure scale are therefore $t=d/U$ and $\rho U^2$, where $\rho$ is the density of the fluid. The equations are solved with the  boundary conditions:
\begin{align}
\textit{\textbf{v}}(\bm{{x}},t) &= 2(1-4r^{2}) \bm e_{z}  \ \ \ \ \ \ \ \ \bm x \in Inlet \label{eqn:FHP} \, , \\
\textit{\textbf{v}}(\bm{x},t) &=  0  \label{eqn:noslip} \ \ \ \ \ \ \ \ \ \ \ \ \ \ \ \ \ \ \ \ \ \   \ \bm x \in Wall, \\
P\bm n - \bm n \cdot \nabla \textit{\textbf{v}}(\bm x,t)/Re &= 0 \ \ \ \ \ \ \ \ \ \ \ \ \ \ \ \ \ \ \ \ \ \  \ \bm x \in Outlet, \ \label{eqn:out}  
\end{align}
corresponding to a fully developed Hagen-Poiseuille flow (\ref{eqn:FHP}) at the inlet, no-slip (\ref{eqn:noslip}) at the walls, and a open boundary condition (\ref{eqn:out}) at the outlet of the pipe. The equation (\ref{eqn:out}) is a Neumann boundary at the outlet, with $\bm n$ being the surface vector pointing outwards from the computational domain, chosen to avoid numerical oscillations. The initial condition used here was a parabolic velocity profile within the inlet pipe section as well as in the outlet section. The velocity jump, near the expansion, adjusts within few time steps. Nek5000 \cite{Fischer2008}, an open source code, has been used to solve the above equations. Spectral elements using Lagrange polynomials are used for spatial discretisation of the computational domain. The weak form of the equation is discretised in space by Galerkin approximation. $N^{\rm th}$ order Lagrange polynomial interpolants on a Gauss-Lobatto-Legendre mesh were chosen as the basis for the velocity space, similarly for the pressure space. The viscous term of the Navier-Stokes equations are treated implicitly using third order backward differentiation and the non-linear terms are treated by a third order extrapolation scheme making it semi-implicit. The velocity and pressure were solved with same order of polynomial. 

\begin{figure}
    \centering
    \includegraphics[width=1.0\textwidth]{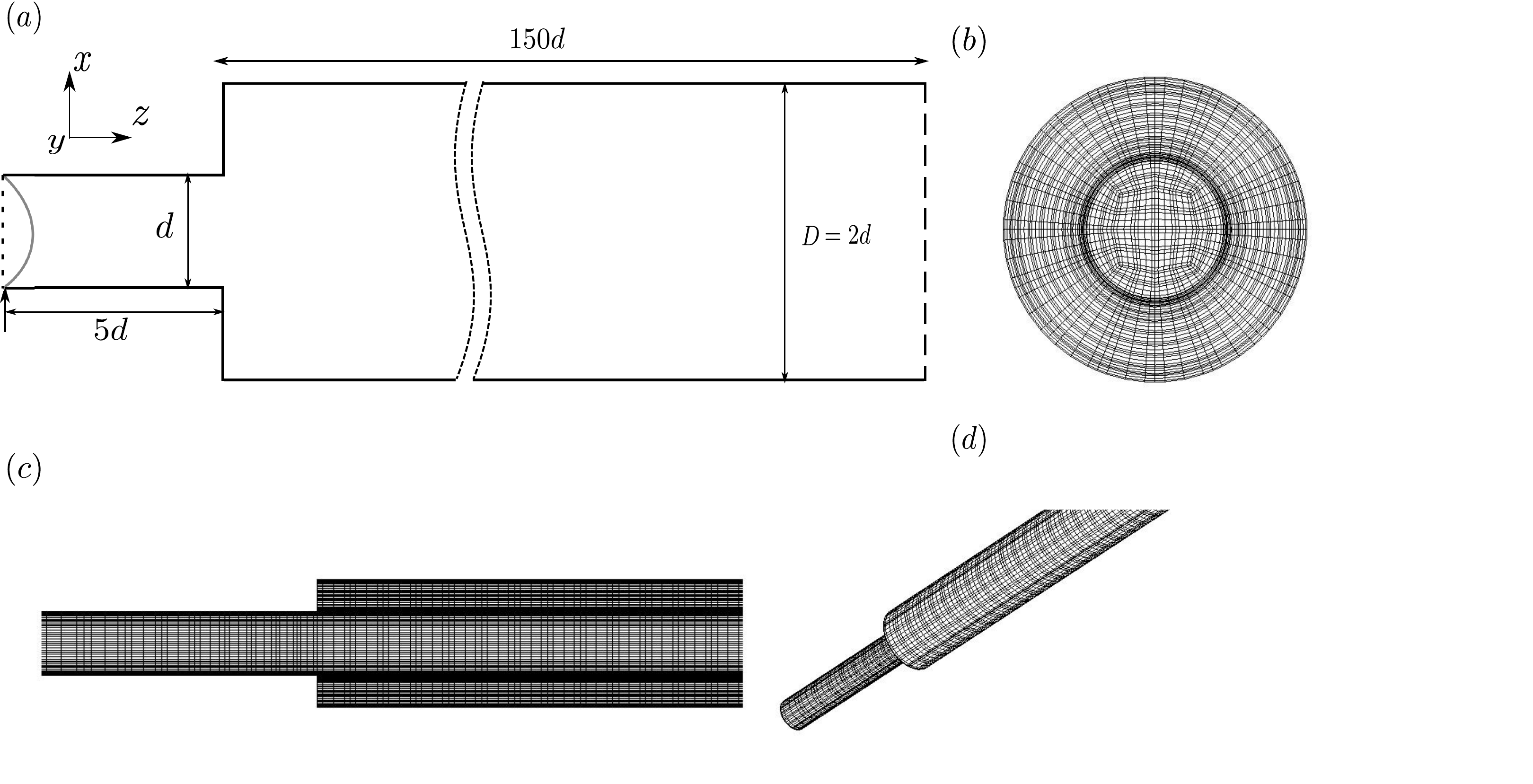}
    \caption{The spectral-element mesh of the sudden expansion pipe. ({\it a}) Sketch of the domain, ({\it b}) $(x,y)$ cross-section of the mesh (the dark lines represent the elements and the grey lines represent the Gauss-Lobatto-Legendre mesh), ({\it c}) $(x,z)$  cross section of the pipe around the expansion and ({\it d}) truncated three dimensional view of the expansion pipe. The mesh is made of $K= 63,200$ elements.}
    \label{fig1}
\end{figure}

Figure \ref{fig1}({\it a}) is a schematic diagram of the expansion pipe. The length of the inlet pipe is $5d$, the outlet pipe is $150d$, and the expansion ratio is given by $E=D/d=2$, where $D$ is the outlet pipe diameter. The computational mesh was created using hexahedral elements. Figure \ref{fig1}$(b)$ shows the $(x,y)$ cross section of the pipe with 160 elements and the streamwise extent of the pipe has 395 elements. The mesh is refined near to the wall and near the expansion section (see figure \ref{fig1}({\it c})). A  three dimensional view of the mesh along the expansion pipe is displayed in figure \ref{fig1}({\it d}). The mesh used here contains approximately four times more elements than our previous study \cite{Selvam2015}. Table \ref{table1} shows the parameters used to assess convergence: (i) the flow reattachment point, $z_r$, and (ii) the viscous drag. The convergence study was done at $Re=1000$  ($z_r$ is very sensitive and may be affected by the outlet at larger $Re$) and no qualitative changes were found for $Re=2000$. $N=5$ is sufficient to resolve the flow accurately near the separation point as well as at the reattachment point. The total number of grid points in the mesh is approximately $KN^{3}=7.9 \times 10^6$, where $K$ is the number of elements. The entire set of simulations reported here took over one calendar year to complete on four processors.

\begin{table}
\begin{center}
    \begin{tabular}{cccc}
    \hline
    \hline
     $N$   &  $KN^3~(\times10^6)$   &  Reattachment Position $z_{r}~$     &  Viscous Drag         \\ \hline
    4        &     4.0       &    43.58 &    0.3725     \\ 
    5        &     7.9       &    43.72  &    0.3333     \\  
    6        &     13.6       &    43.73  &    0.3323     \\  
    \hline
    \hline
    \end{tabular} 
   \caption{Convergence study, changing the order of polynomial $N$. $z_{r}$ is the non-dimensional length of the recirculation region in the pipe for $Re=1000$.} 
    \label{table1}
\end{center}
\end{table}

\section{Vortex perturbation, effect of the amplitude of the vortex perturbation and proper orthogonal decomposition}

\subsection*{Vortex perturbation}
When trying to make connection between experimental observations and simulations, the issue of the choice of perturbation must be addressed. Many perturbations have been tested experimentally \cite{Darbyshire1995,Peixinho2007,Nishi2008,Mullin2011} and replications in numerical works have reproduced some of the observations \cite{Mellibovsky2007,Asen2010,Loiseau2014b,Wu2015}. 

\begin{figure}
    \centering
    \includegraphics[width=1.0\textwidth]{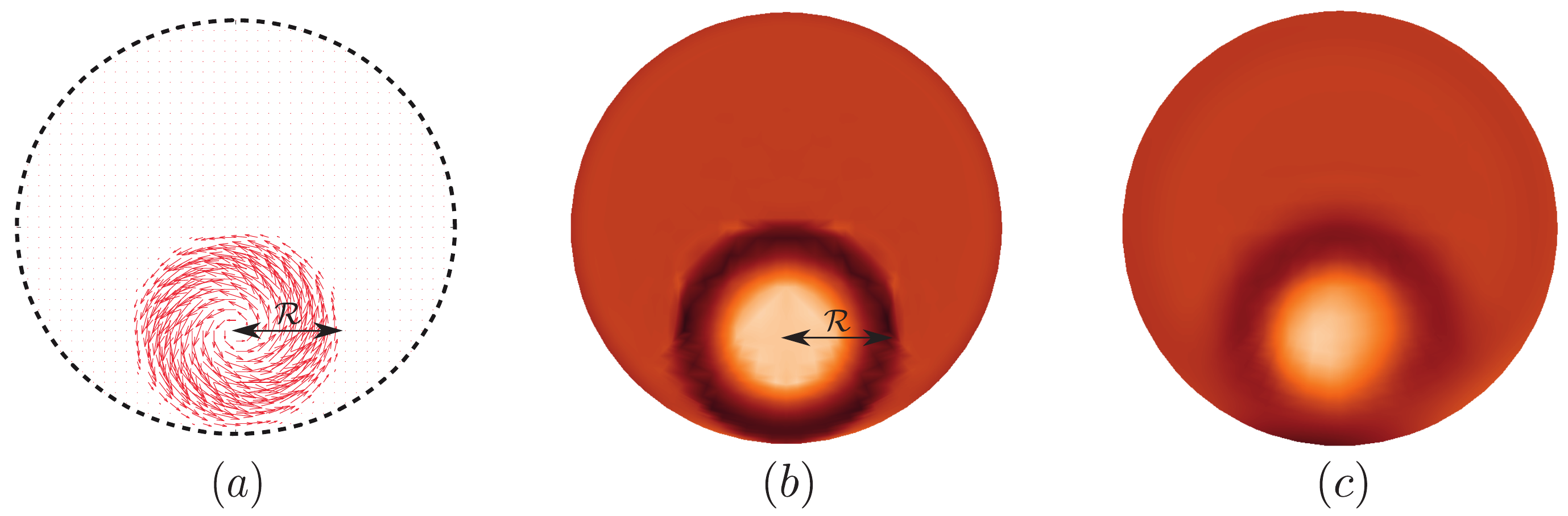}
    \caption{({\it a}) Vector plot of $\vec{u}'$.  Axial vorticity contour of the vortex perturbation ($\mathcal{R}=0.25$) in the inlet of the pipe at ({\it b}) $z=-5$ and ({\it c}) $z=-2.5$ for $Re= 2000$. Black and white corresponds to the maximum and minimum of vorticity and orange (grey) represents zero vorticity.}
    \label{fig2}
\end{figure}

Here, we aim to consider a simple localized perturbation, and introduce
a localized vortex to the inlet Poiseuille flow. The radial size of the vortex may be controlled as well as its position in the inlet section. This perturbation also satisfies the continuity condition at the injection point and automatically breaks axisymmetry, contrary to the tilt perturbation \cite{Sanmiguel-Rojas2010,Selvam2015}

We define $s=\sqrt{ (x-x_0)^2 + (y-y_0)^2 }$ as the distance between the center of the vortex at $(x_0,y_0)$ to any point $(x,y)$ in the cross-section, at which the local measure of rotation is given by
\begin{eqnarray}
\Omega & = & \left\{ \begin{array}{ll}
      1,        & s \le \mathcal{R}/2, \\
      2(\mathcal{R}-s)/R, & \mathcal{R}/2 < s \le \mathcal{R} , \\
      0,        & s > \mathcal{R} \,,
   \end{array}\right.
\end{eqnarray}
where $\mathcal{R}$ is the radius of the vortex.  The
velocity perturbation $\vec{u}'$ in Cartesian coordinates is then
\begin{equation}
    \vec{u}' =  \delta \, \Omega \, ( y_0-y,\, x-x_0,\, 0 ) \, ,
\end{equation}   
where $\delta$ is a parameter measuring the strength of the vortex. 
The full inlet condition is therefore
\begin{eqnarray}
 \vec{u} & = & \vec{U} + \vec{u}'\, , \\ \nonumber
    & = & ( 0,\,0,\,U(r)) ~ +  ~
    \delta \, \Omega \, ( y_0-y,\, x-x_0,\, 0 ) \, , \\  
    & = & ( \delta\,\Omega(y_0-y),\, \delta\,\Omega(x-x_0), \,U(r) )  \, .
\end{eqnarray}
The parameter $\mathcal{R}=0.25$ is kept constant in all the present simulations. The perturbation is added at the inlet pipe along with the parabolic flow velocity profile at $z=-5$. Figure \ref{fig2}({\it a}) is a cross-section of velocity field of the vortex perturbation. Figure \ref{fig2}({\it b}) and ({\it c}) show contour plots of axial vorticity at the inlet section of the pipe, $z=-5$, and further downstream at $z=-2.5$. The contours show that the perturbation diffuses and becomes smoother along the inlet. At the expansion section, $z=0$, perturbations are known to be amplified \cite{Cantwell2010}.

\subsection*{Effect of amplitude of the vortex perturbation}

\begin{figure}[htp]
    \centering
    \includegraphics[width=1\textwidth]{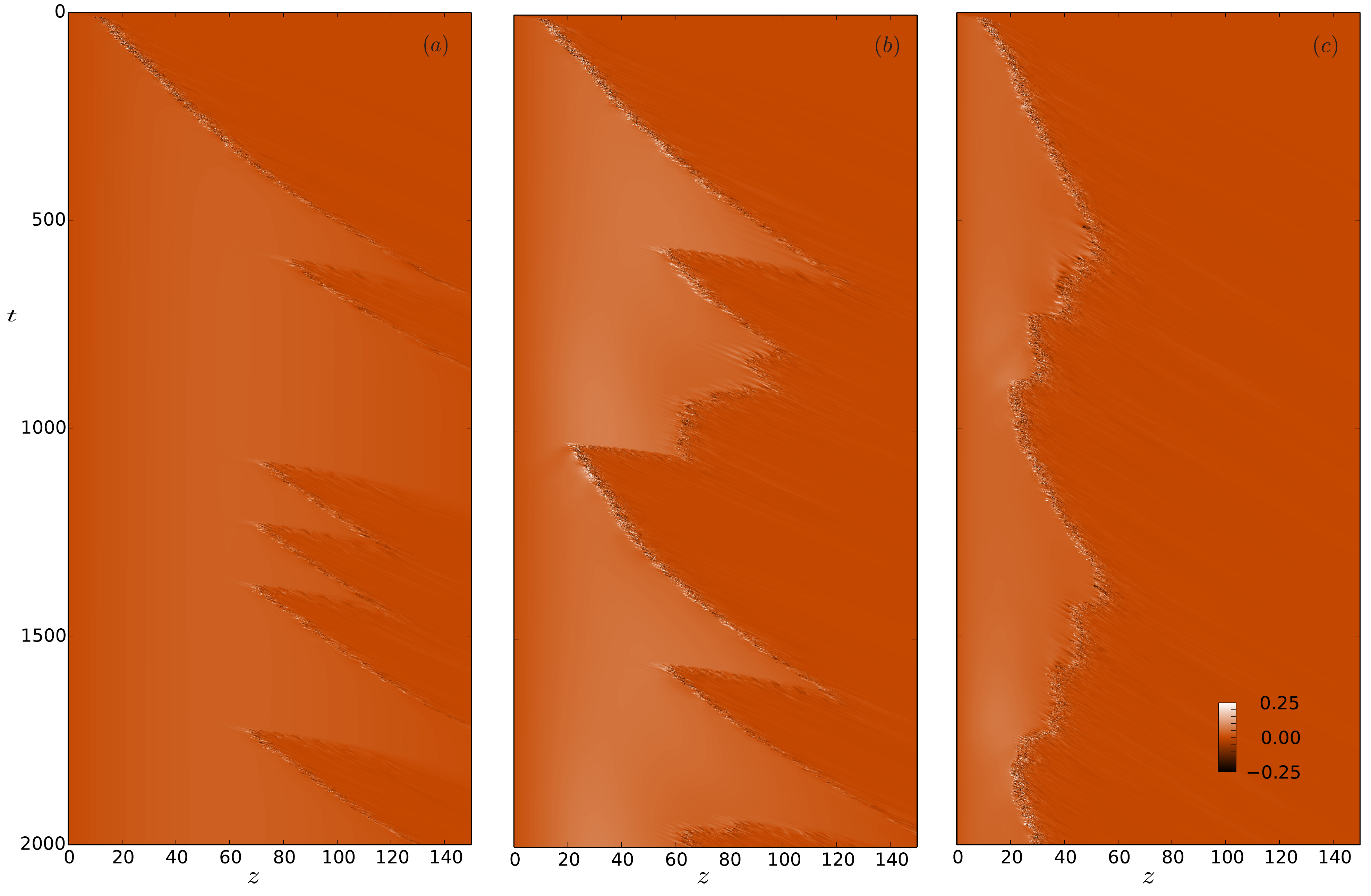}
    \caption{Spacetime diagram of the centreline streamwise vorticity for $Re=2000$ for ({\it a}) $\delta=0.05$, ({\it b})  $\delta=0.1$, and ({\it c})  $\delta=0.2$.}
    \label{fig3}
\end{figure}

In previous works \cite{Sanmiguel-Rojas2010,Selvam2015}, the addition of a tilt perturbation has been found to trigger transition to turbulence. However, the tilt perturbation (i) creates a discontinuity at the inlet and (ii) does not break the mirror symmetry. In this respect, the vortex perturbation permits a more controlled transition, resulting in smoother dependence of the transitional regime on the strength of the perturbation. Figure \ref{fig3} shows a space-time diagram for the centreline streamwise vorticity at $Re=2000$ for different perturbation strengths, $\delta$. After $t\approx500$, it can be seen that for different $\delta$ the flow settles into different behaviours of the turbulent patches, observed over the following 1500 time units.  Computational costs limit
simulations to larger $t$.

For $\delta<0.05$, the perturbation decays before reaching the expansion section. At $\delta=0.05$ (see figure \ref{fig3}({\it a})), a turbulent localized patch forms, then moves downstream. Around $t\simeq600$ another turbulent patch forms upstream at $z\simeq60$ and the downstream patch decays immediately. This process appears to repeat in a quasi-periodic manner.

When the amplitude of the vortex perturbation is increased, $\delta=0.1$, see figure \ref{fig3}({\it b}), again a patch of turbulence appears, then moves downstream. When a turbulent patch arises upstream at $t\simeq600$, the patch  downstream again decays immediately. This time, however, the process appears to repeat more stochastically, in time and location, of the arising upstream patch. Occasional reversal in the drift of the patch is also observed. It is expected that if the patch drifts far downstream, then it will relaminarise, since the the local Reynolds number based on the outlet diameter is $Re/E=1000$, somewhat below the $2000$ typically required for sustained turbulence. It is likely that the deformation to the flow profile by the  upstream patch reduces the potential for growth of perturbations within the patch downstream, disrupting the self-sustaining process.

\begin{figure}[htp]
    \centering
    \includegraphics[width=1\textwidth]{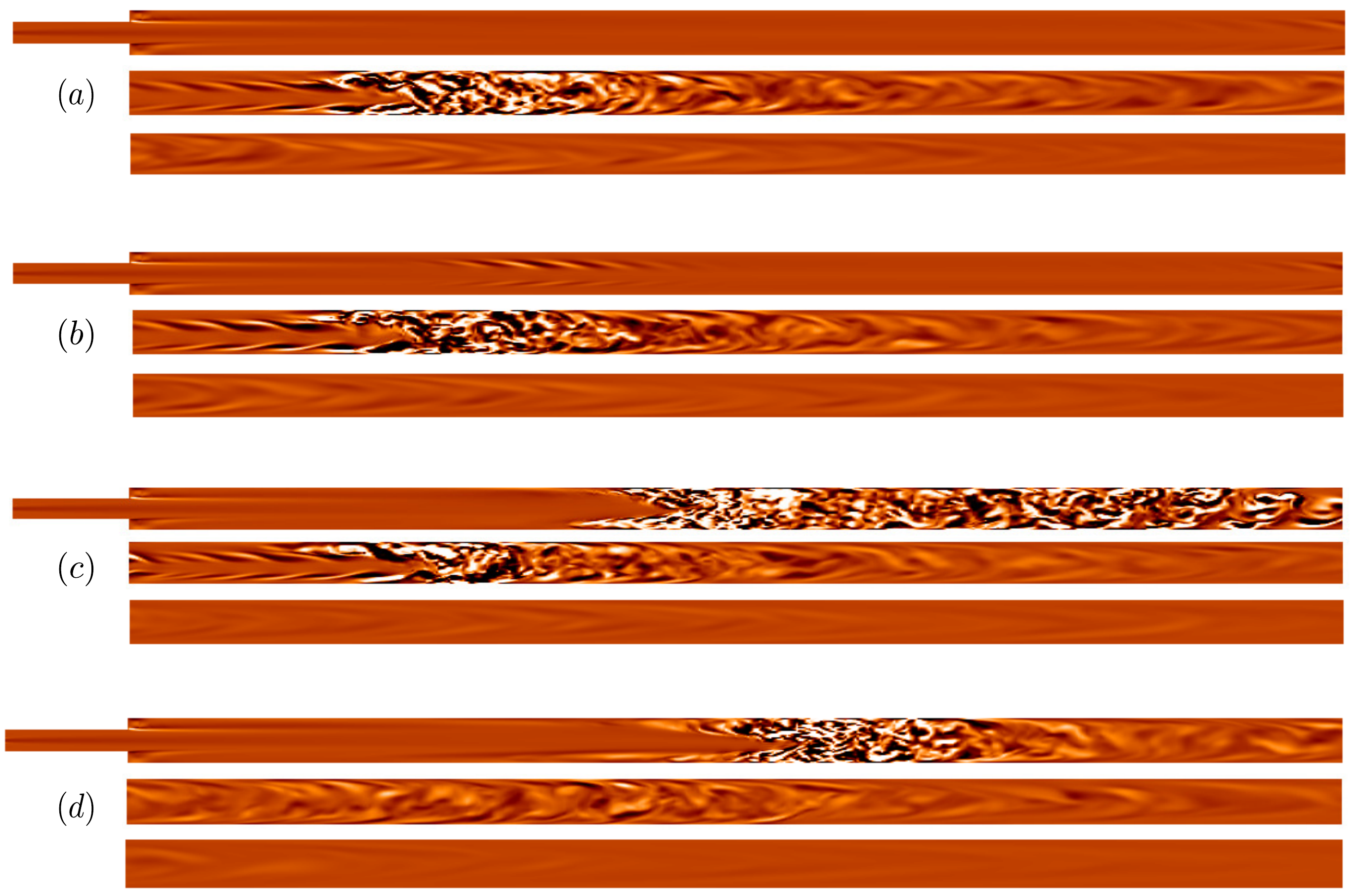}
    \caption{ $x-z$ cross sections of streamwise vorticity contour plot for $Re$ = 2000 with $\delta = 0.1$ at $(a)$ $t = 1000 $, $(b)$  $t=1025$, $(c)$ $t=1050$ and $(d)$ $t=1100$.  Each triad represents the full pipe length, truncated  at every $50d$ for simple visualization purpose. Here black and white corresponds to the maximum and minimum of vorticity and orange (grey) represents zero vorticity.}
    \label{fig4}
\end{figure}

\begin{figure}[htp]
    \centering
    \includegraphics[width=0.5\textwidth]{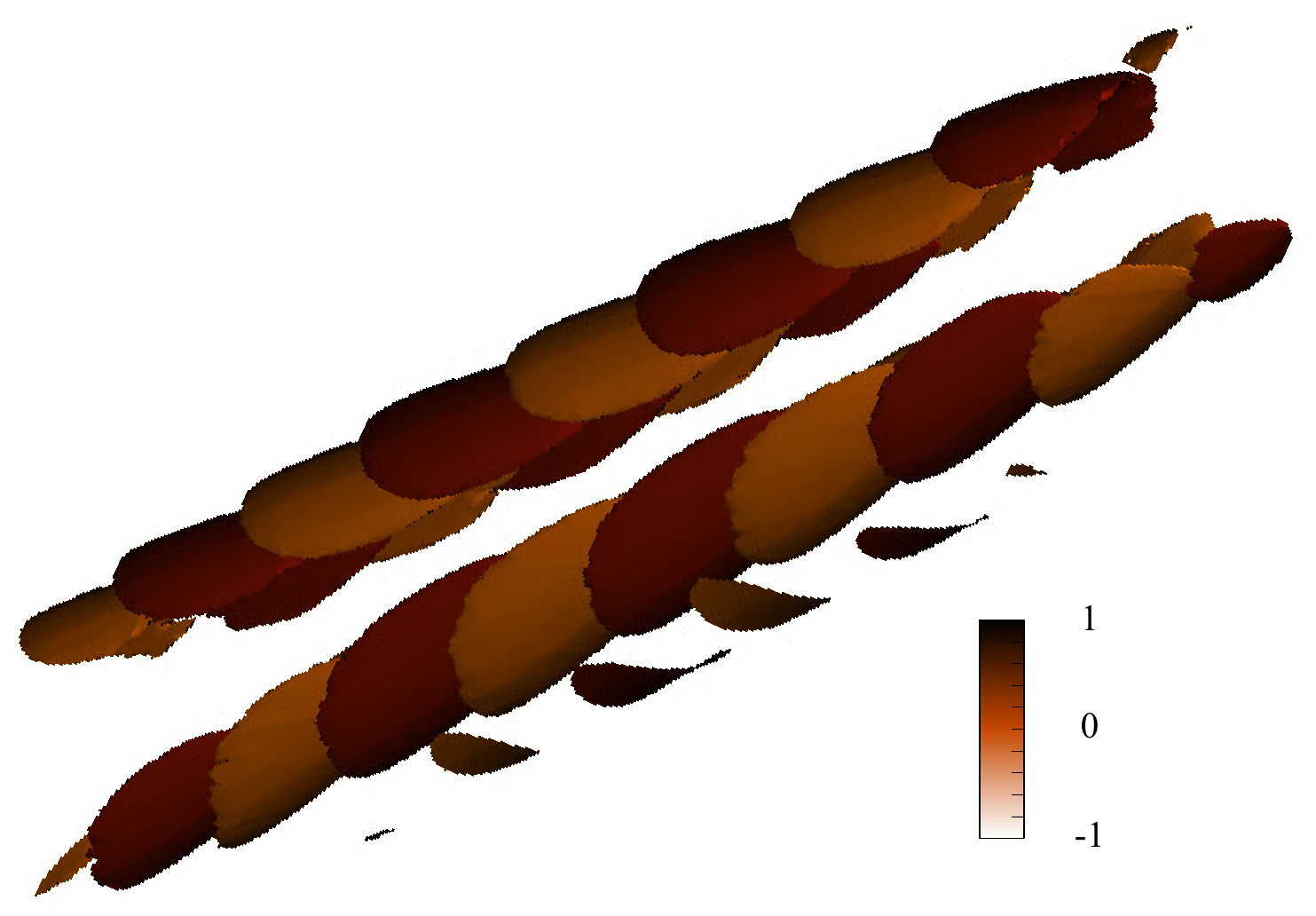}
    \caption{Iso-surface of streamwise vorticity resembling the optimal perturbation for $Re=2000 $, $ \delta=0.1$ at $t=1025$ and spanning from $z=12.5$ to 25 from left to right.}
    \label{fig5}
\end{figure}

Still for $\delta=0.1$, figure \ref{fig4} shows the streamwise vorticity for a $(x,z)$ cross-section over the whole pipe: $150d$.  At $t=1000$ (see figure \ref{fig4}({\it a})), it can be seen that only a single turbulent patch exists in the domain. At $t = 1025$ (see figure \ref{fig4}({\it b})), an axially periodic structure appears at $z\simeq10$. Once this develops into turbulence (see figure \ref{fig4}({\it c})), the patch downstream dissipates rapidly (see figure \ref{fig4}({\it d})). The appearance of the new patch in our expansion is different from the puff splitting process observed in a straight pipe \cite{Wygnanski1973,Nishi2008,Duguet2010,Moxey2010,Hof2010,Avila2011,Shimizu2014,Barkley2015}. Here the new turbulent patch evolves out of the amplified perturbation at the entrance and breaks down into turbulence, forming a new patch upstream of an existing patch. The patch drifts downstream and decays. The slopes in the diagrams of figure \ref{fig3} indicate the drift velocity of the patch, which varies with respect to $\delta$ and $z$, and decreases as $\delta$ increases. Figure \ref{fig5} shows the iso-surface streamwise vorticity for the axially periodic structure that appears at $z\simeq10$, in this case it is shown for $12.5<z<25$ at $t=2000$. This structure appears repeatedly and resembles the optimally amplified perturbation found in a sudden expansion flow by \cite{Cantwell2010}.  Initially the structure appears near the expansion region, where the flow is very sensitive to perturbations, it is amplified and then breaks down into turbulence downstream.

For $\delta=0.2$, see figure \ref{fig3}({\it c}), the turbulent patch never goes beyond $z\simeq60$.  Here the perturbation develops consistently into turbulence, so that its position remains roughly constant. The patch remains close enough to the entrance so that there is insufficient space for a new distinct patch to arise. 

\begin{figure}[htp]
    \centering
    \includegraphics[width=1\textwidth]{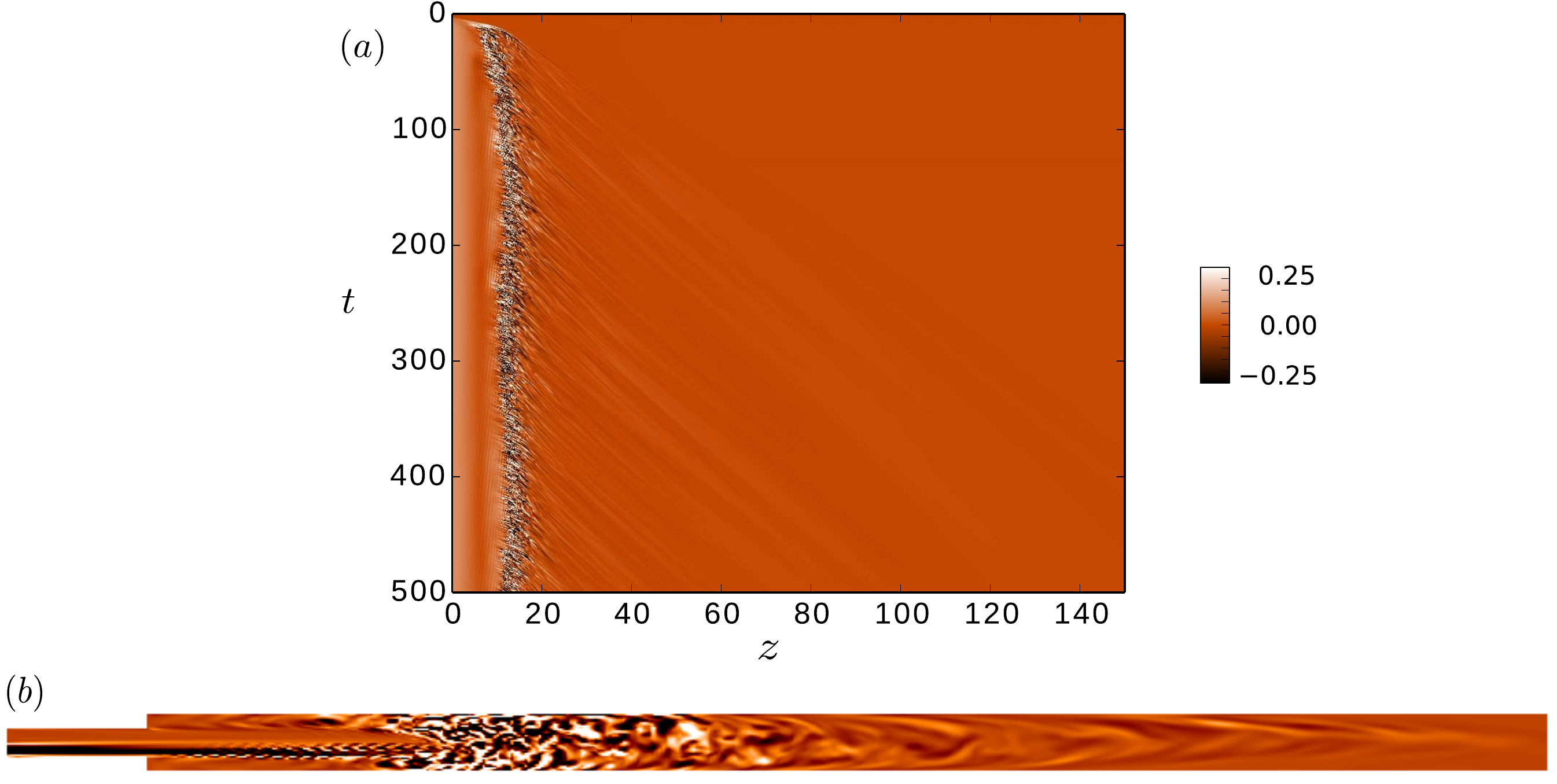}
    \caption{({\it a}) Spacetime diagram for the centreline streamwise vorticity for $Re=2000$ and $\delta=0.5$. ({\it b}) Zoomed contour plot of the streamwise vorticity for $z$ up to 50, black and white corresponds to the maximum and minimum of vorticity and orange (grey) represents zero vorticity. Note the perturbation development between the expansion section and the turbulent patch.}
    \label{fig6}
\end{figure}

For large amplitude $\delta=0.5$, the turbulence patch does not drift, remaining at a more stable axial position, shown in the spatiotemporal diagram of figure \ref{fig6}({\it a}). A snapshot of the flow at $t=100$ is also presented in figure \ref{fig6}({\it b}), and this streamwise vorticity contour plot highlights the effect of the vortex perturbation that is clearly at the origin of the turbulent patch. 

In previous works \cite{Sanmiguel-Rojas2010,Selvam2015}, spatially localized turbulence has also been observed, and one question that can be asked is how similar or different is this localized turbulence from the turbulent puffs observed in straight pipe flow \cite{Wygnanski1973}? Using spatial correlation functions, previous works \cite{Selvam2015} have found that the localized turbulence in expansion pipe flow is more active in the centre region than near the wall, hence different from the puffs in uniform pipe flow \cite{Willis2008}. In the next section, we provide results on a another analysis tool: the proper orthogonal decomposition.

\subsection*{Proper Orthogonal Decomposition of turbulence }

Principle Component Analysis, often called Proper Orthogonal Decomposition (POD) in the context of fluid flow analysis, has been widely used by several researchers \cite{Lumley1967,Noack2003,Sirovich1987,Meyer2007} to identify coherent structures in turbulent flows by extracting an orthogonal set of principle components in a given set of data. Each data sample $a_i$, being a snapshot state, may be considered as a vector in $m$-dimensional space, where $m$ is e.g.~the number of grid points. These vectors may be combined to form the columns of the $m\times n$ data matrix $\bm{X}=[a_1\,a_2\,\dots\,a_n]$, where, $n$ is the number of snapshots. Let $\bm{T}$ be an $m \times n $ matrix with columns of principle components, related by to $\bm{X}$ by
\begin{equation}
\bm{T} = \bm{X} \bm{W} \,.
\end{equation}
$\bm{T}$ is intended to be an alternative representation for the data, having columns of orthogonal vectors with the property that the first $n'$ columns of $\bm{T}$  span the data in $\bm{X}$ with minimal residual, for any $n'<n$. Here the inner product $a^Ta$ corresponds to the energy norm for  the minimisation.

$\bm{W}$ is defined via the singular value decomposition (SVD) of the covariance matrix $\bm{X}^T \bm{X}$. If the SVD of $\bm{X}$ is
  \begin{equation}
  \bm{X}  =  \bm{\tilde{U}} \Sigma \bm{W}^{T} \, ,
  \end{equation}
where, $\Sigma $ is the diagonal matrix of the singular values, then
  \begin{equation}
  \bm{X}^T \bm{X}  =  \bm{W}\Sigma^{T}\bm{\tilde{U}}^{T} \bm{\tilde{U}} \Sigma \bm{W}^{T} 
  = \bm{W}\Sigma^2\bm{W}^T \, .
 \label{svdx} 
 \end{equation}
Also the SVD of $\bm{X}^T\bm{X}$ may be calculated,
  \begin{equation}
  \bm{X}^T \bm{X}  =   \bm{U} \bm{S} \bm{V}^{T} \, .
  \label{svdxx}
  \end{equation}
Comparing equation (\ref{svdx}) and (\ref{svdxx}) we have that $\bm{W}  \equiv \bm{U} .$ Therefore, to calculate the principle components we construct the $n\times n$ matrix of inner products $\bm{X}^T \bm{X}$, where it is assumed that $n\ll m$, and compute its SVD (\ref{svdxx}). Only the first columns of $\bm{T}$ are expected to be of interest, and the $j^\mathrm{th}$ principle component $\hat{u}_j$ may be obtained by
\begin{equation}
u_{j} = \sum_{i=1}^n a_{i} \bm{U}_{i,j},
\qquad
\hat{u}_j = u_{j} / (u^T_{j} u_{j}) .
\end{equation} 
The normalised singular values 
\begin{equation}
\hat{\Sigma}_{jj} = \sqrt{\bm{S}_{jj}/(n-1)},
\end{equation} 
are a measure of the energy captured by each component, having the property that $\hat{\Sigma}_{jj}$ equals the root mean square of $a_i^T\hat{u}_j$ over the data set.

\begin{figure}[htp]
    \centering
    \includegraphics[width=1\textwidth]{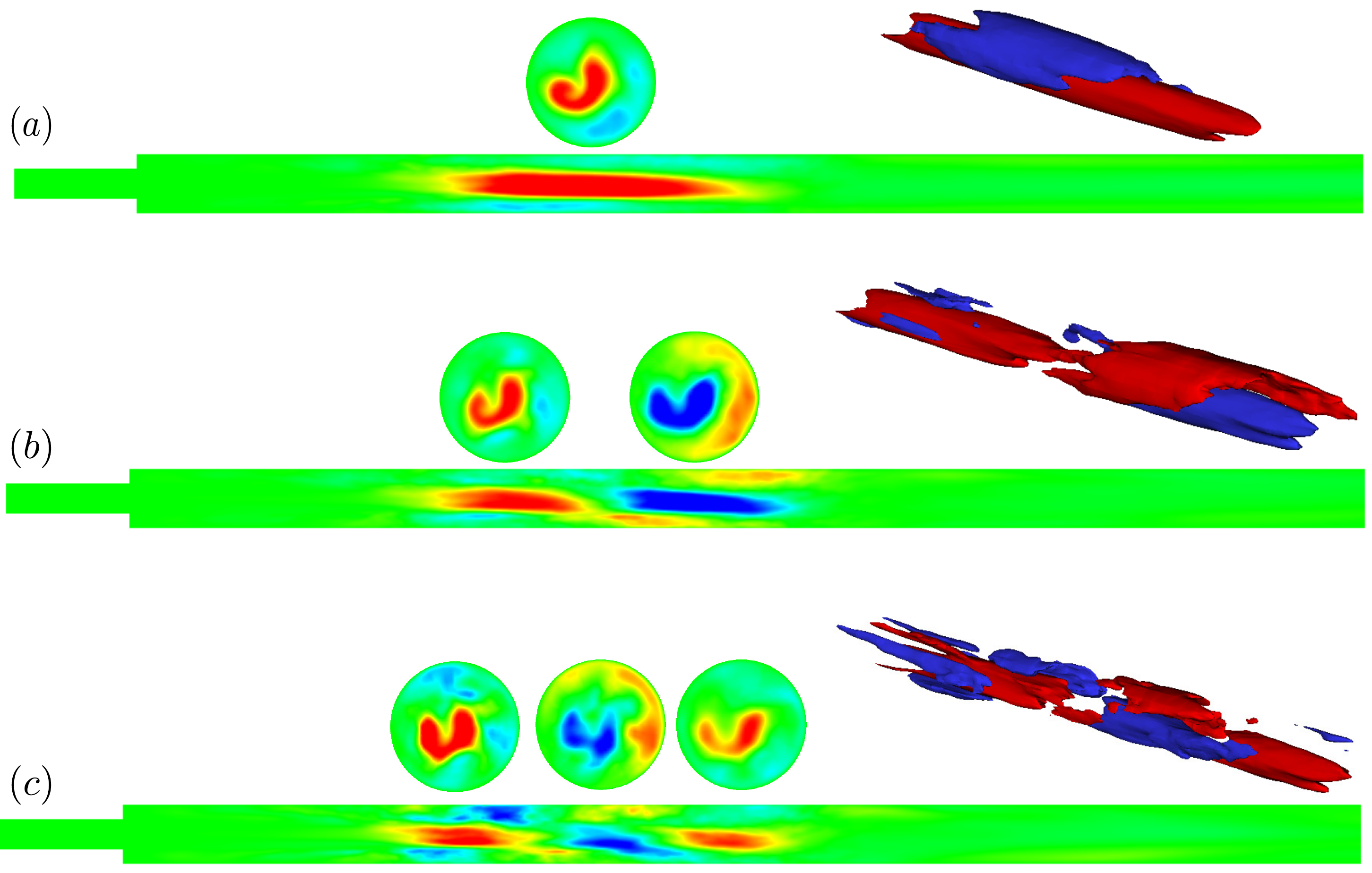}
    \caption{Cross sections ($x,z$), ($x,y$) and iso-surfaces of the proper orthogonal decomposition.  ({\it a}) Mode 1, ({\it b}) mode 2 and  ({\it c}) mode 3 computed for $Re=2000$ and $\delta=0.5$ using 1500 snapshots. Red (light-gray) and blue (dark-gray) correspond to the maximum and minimum of streamwise velocity component.}
    \label{fig7}
\end{figure}

A large number of snapshots were collected, and it was been found that after 1200 snapshots the energy of the leading POD modes (principle components) became independent of the number of snapshots. Figure 7({\it a}) shows the axial velocity of mode 1, which constitutes 74\% of the total kinetic energy. It can be seen that the centre core region is predominant and its shape is reminiscent of the vortex perturbation. Hence, the inlet flow has more effect on the localised turbulence than the wall shear. Mode 2 is shown in the figure 7({\it b}), has two predominant region along the axial direction and constitutes $\approx20$\% of the energy. Mode 3 represents only $\approx3$\% of the energy and is shown in the figure 7({\it c}). The remaining modes appear more complex and less energetic.

In addition, simulations were carried out by changing $\mathcal{R}$ and $(x_0,y_0)$ independently. It has been found that (i) a smaller vortex perturbation: $\mathcal{R}\lesssim0.2$  and (ii) a vortex closer to the centreline could not sustain a fixed localized turbulent patch \cite{Wu2015}.

\section{Conclusions}

Numerical results for the flow through a circular pipe with a sudden expansion in presence of a vortex perturbation at the inlet have been presented.  For $Re=2000$ and a relatively small perturbation amplitude, $0.05\lesssim\delta\lesssim0.1$, a patch of turbulence in the outlet section is observed to drift downstream, then decay upon the appearance of another patch of turbulence upstream. Moreover, this vortex perturbation produces a controlled transition, in that the transitional regime depends smoothly on the perturbation strength, and the origin of symmetry breaking is defined. Further, the turbulent patch that forms first appears via a low order azimuthal mode resembling an optimal perturbation. The process repeats quasi-periodically or stochastically as the amplitude of the perturbation, $\delta$, increases. The turbulent patch formation is different from the puff splitting behaviour observed in uniform pipe flow \cite{Wygnanski1973,Hof2010,Avila2011,Barkley2015}, as here the new patches arise upstream of existing turbulent patches.

The drift velocity of the patch varies with $\delta$, decreasing as $\delta$ is increased. For large $\delta$, the patch does not drift downstream, but holds a stable spatial position forming localized turbulence. The structure within the localised turbulence is further studied using proper orthogonal decomposition, which indicates that the first mode comprises most of the energy and the flow is more active in the centre region than near the wall. 

\section*{Acknowledgements}
This research was supported by the {\it R\'{e}gion Haute Normandie}, the {\it Agence Nationale de  la Recherche} (ANR)  and the computational time provided by the {\it Centre R\'egional Informatique et d'Applications Num\'eriques de Normandie} (CRIANN). Our work has also benefited from many discussions with D. Barkley,  J.-C. Loiseau  and T. Mullin. 

\section*{References}
\bibliography{DivergentPipeBib}
\bibliographystyle{jphysicsB}
\end{document}